\documentclass[aps,prl,twocolumn,floatfix]{revtex4}

\usepackage{hyperref}
\usepackage{graphicx}
\usepackage[utf8]{inputenc}
\usepackage{amsmath}
\usepackage{comment}
\usepackage{bm}
\usepackage{dsfont}
\usepackage{amsfonts}
\usepackage{xcolor}

\begin{document}

\title{Probing spin fractionalization with ESR-STM absolute magnetometry}

\author{Y. del Castillo$^{1,2}$, J. Fern\'{a}ndez-Rossier$^1$\footnote{On permanent leave from Departamento de F\'{i}sica Aplicada, Universidad de Alicante, 03690 San Vicente del Raspeig, Spain}$^,$\footnote{joaquin.fernandez-rossier@inl.int} }

\affiliation{$^1$International Iberian Nanotechnology Laboratory (INL), Av. Mestre Jos\'{e} Veiga, 4715-330 Braga, Portugal }
\affiliation{$^2$Centro de F\'{i}sica das Universidades do Minho e do Porto, Universidade do Minho, Campus de Gualtar, 4710-057 Braga, Portugal }
\date{\today}


\begin{abstract}
The emergence of effective $S=1/2$ spins at the edges of $S=1$ Haldane
spin chains is one of the simplest examples of fractionalization.  
Whereas there is indirect evidence of this phenomenon, 
 direct measurement of the magnetic moment of an individual edge spin remains to be done. Here we show how scanning tunnel microscopy electron-spin resonance (ESR-STM) can be used to  map the stray field created by the fractional  $S=1/2$
edge spin and we propose
efficient methods to invert the Biot-Savart equation,  obtaining the edge magnetization map. This permits one to determine unambiguously 
the two outstanding emergent properties of fractional degrees of freedom, namely, their fractional magnetic moment and their localization length $\xi$.
\end{abstract}

\maketitle

Fractionalization is one of the most dramatic examples of emergence in many-body systems \cite{su80,tsui82,laughlin83}.  It shows how new quantized degrees of freedom, such as quasiparticles with charge $e/3$, can govern the low-energy properties of a system of interacting electrons with charge $e$. In the case discussed here,  a chain of interacting $S=1$ spin behaves as if two $S=1/2$ spin degrees of freedom were localized at the edges. These examples illustrate how we can not rule out that the quantum numbers of the so-called fundamental particles are actually emerging out of an interacting system made of degrees of freedom with different quantum numbers \cite{laughlin00}. 

Haldane spin chains \cite{haldane83a,haldane83b} provide one of the simplest examples of fractionalization and emergence. Out of a model of interacting $S=1$ spins without intrinsic energy and length scales, a Haldane gap $\Delta_H$, and two $S=1/2$ degrees of freedom localized at the edges with localization length $\xi$ emerge. Given that the building blocks of the model have $S=1$,  the $S=1/2$ edge states are fractional. Their emergence can be rationalized in terms of the AKLT \cite{Affleck87} valence bond solid state, that in turn has a number of outstanding properties, including being a resource state for measurement-based quantum computing \cite{Wei2011}. 

The fractional charge of quasiparticles in the Fractional Quantum Hall effect was determined by an outstanding experiment\cite{saminadayar97,de98} that leveraged on the relation between shot noise and charge\cite{kane94}. In the case of spin fractionalization, a direct measurement of the spin of the fractional edge states and their localization length remains to be done. Until recently, experimental probes of Haldane spin chains relied on bulk probes, such as neutron scattering \cite{buyers86,tun90,yokoo95,zaliznyak01,kenzelmann03} and electron spin resonance \cite{brunel92,batista99,smirnov08}, and provided indirect evidence of the presence of $S=1/2$ degrees of freedom and a Haldane gap. Advances in on-surface synthesis combined with atomic-scale resolution inelastic electron tunnel spectroscopy (IETS) based on scanning tunnel microscopy (STM) have made it possible to probe individual Haldane spin chains made with covalently coupled $S=1$ nanographene triangulenes \cite{Mishra2021}. 
IETS of triangulene spin chains showed the presence of a Haldane gap in the center of the chains as well as in-gap edge excitations for short chains and zero bias Kondo peaks for longer chains \cite{Mishra2021},  consistent with the existence of emergent $S=1/2$ edge spins \cite{delgado13}. 

Here we propose a {\em direct} measurement of the edge magnetic moment $M=g\mu_B S_{\rm edge}$ associated to the fractional  $S_{\rm edge}=1/2$ degrees of freedom using STM-based electron spin resonance (ESR-STM)\cite{baumann15}. We assume that a Haldane spin chain, not necessarily made with triangulenes, is deposited on a surface, sufficiently decoupled from the substrate so that the Kondo effect is suppressed, the magnetic moment of the edge states is preserved and can be probed with ESR-STM magnetometry \cite{choi17}. The feasibility of this weak-coupling scenario has been demonstrated 
in several ESR-STM experiments where $S=1/2$ species, such as Ti \cite{choi17}, Cu \cite{yang18} and alkali atoms \cite{kovarik2022}, are deposited on a bilayer of MgO on top of an Ag surface. 
\begin{figure}[!ht]
\centering
\includegraphics[width=1\linewidth]{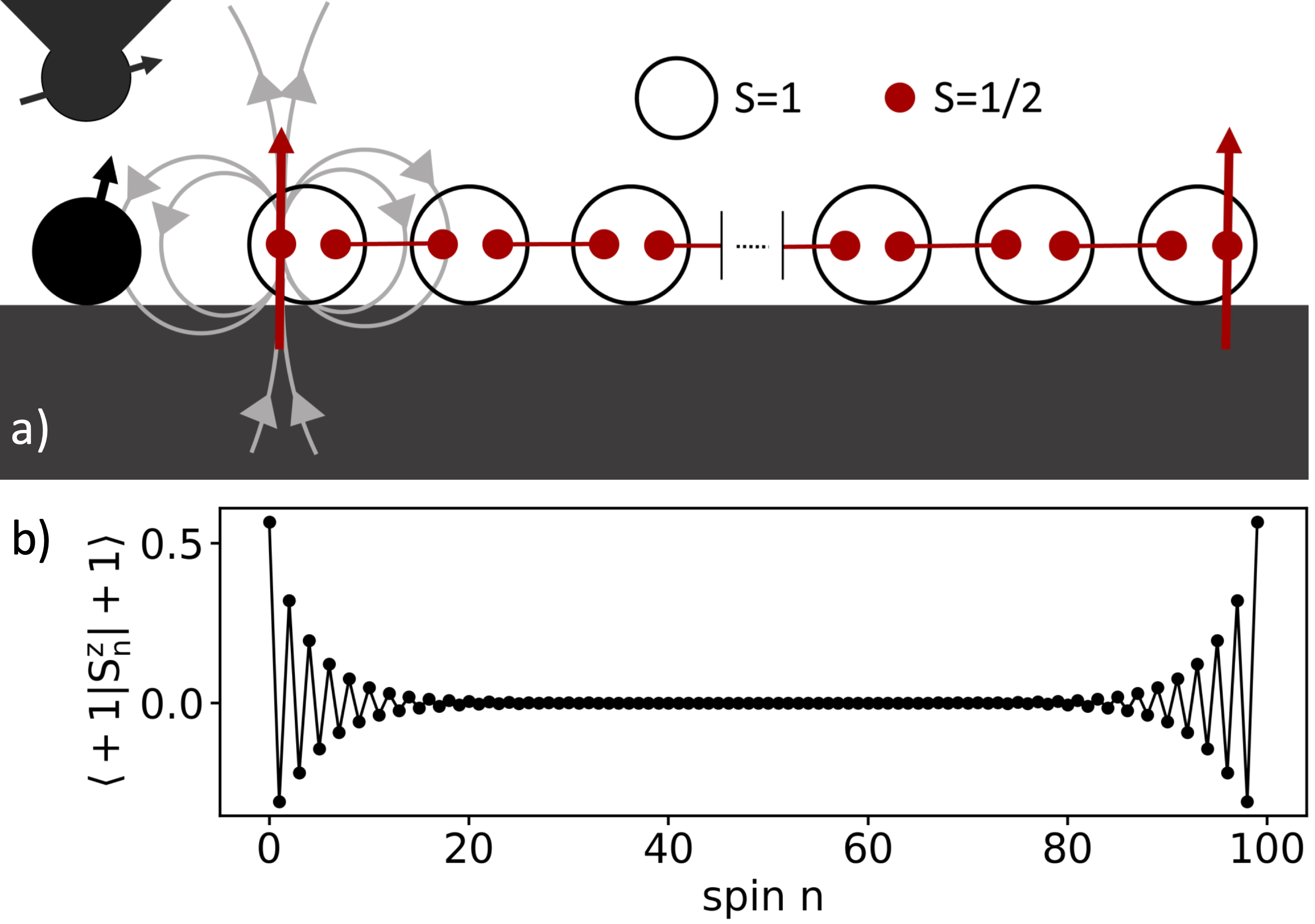}
\caption{a) Scheme of the proposed experiment to measure fractionalization by means of ESR-STM. A Haldane chain is built on a kondo-free surface. The system is described as if there were two $S=1/2$  objects at the edge. An ESR active atom (sensor atom) is placed near the chain to infer the magnetic moments of such chain from its stray field. b) Spin density of the $S^z=+1$ triplet state for $\beta=0.09$.}
\label{fig:1}
\end{figure}

Our proposal (see Fig. \ref{fig:1}) relies on the demonstrated capability of ESR-STM to act as an absolute magnetometer \cite{natterer17,choi17}. To do so, an ESR-STM active spin acts as a sensor that can be placed at several distances of a second spin or group of spins denoted as target. At finite temperature, the target spins can occupy different quantum states, each of which generates its own stray field\cite{choi17,delCastillo23}. As a result, the ESR-STM spectrum of the sensor spin features several peaks, whose frequency and intensity relate to the stray-field and occupation probability of the quantum state of the target. As we show below, this can be used to obtain a {\em direct measurement} of the magnetic moment, and thereby the spin, of the edge states in Haldane spin chains.

We assume that a Haldane spin chain with $N$ spins $S=1$ is placed on a surface and can be described with the Hamiltonian:
\begin{equation}
    \begin{split}
    {\cal H}= \sum_{n=1,N-1} J & \bigg[\vec{S}_n\cdot  \vec{S}_{n+1}+\beta\left(\vec{S}_n\cdot \vec{S}_{n+1} \right)^2 \bigg ] \\
     &+\sum_{n=1,N} g\mu_B \vec{S}_n\cdot\vec{B}
    \end{split}
\label{eq:ham}
\end{equation}
We take values of $\beta$ in the range $0<\beta<\frac{1}{3}$. The low energy manifold is conformed by a singlet with $S=0$ and a $S=1$ triplet. The singlet-triplet splitting is given by the sum of the effective inter-edge coupling and the Zeeman energy:
\begin{equation}
    E(S,S^z)=S \Delta_{ST}(N)+g\mu_B S^z B^z,
\end{equation}
where $\mu_B$ is the Bohr magneton, $g=2$  and $S=0,1$ and $S^z=\pm 1, 0$. We choose the quantization axis of the spin to be parallel to the external magnetic field. 

The singlet-triplet splitting shows an exponential decay given by $\Delta_{ST}\propto e^{-N/\xi}$, where $N$ is the system size and $\xi$ represents the localization length. This quantity remains significantly smaller than the Haldane gap, the energy splitting between the low-energy manifold and the bulk states. 
The exponential dependence of $\Delta_{ST}$ closely resembles what would be expected if two $S=1/2$ spins were localized at the edges of the chain, with a localization length on the order of $\xi$. This characteristic is illustrated in Fig. \ref{fig:1}b, where we calculate the expectation value of the $S^z_n$ operators for the low-energy states with $S=1$ and $S^z=+1$.
Clearly, these states form a magnetic texture localized at each edge, with a combined magnetic moment of $S^z=\sum_{n=1, N/2} \langle \pm 1 | S^z_n| \pm 1 \rangle =  \pm 1/2$. 
\begin{figure*}[t]
\centering
\includegraphics[width=1\textwidth]{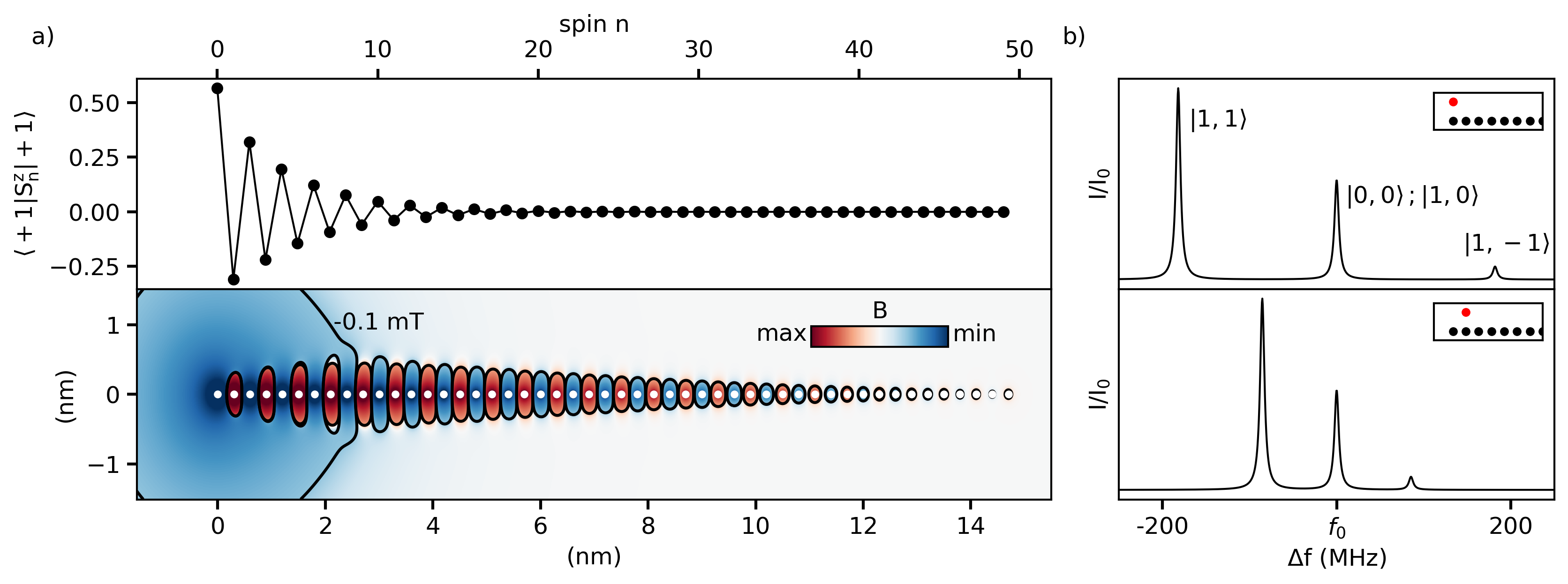}
\caption{ a) The top panel shows the $\langle S^n_z \rangle$ in the $S^z=1$ state of the low-energy manifold for the first half of a Haldane chain of 100 spins with $S=1$ (full chain in the inset). The lower panel illustrates the $z$ component of the stray field generated by the magnetic moments from the spins in the top panel (represented as white dots) on the $xy$ plane. In this example, the chain is located spanning from $(x=0,y=0,z=0)$ to $(x=+Na,y=0,z=0)$. Here, $a$ represents the physical spacing between the $S=1$ spins, treated as point particles. We choose $g=2$ and a spacing of $0.3 \text{nm}$ from each other. The contour lines highlight regions where the stray field has a magnitude of $\pm0.1  \text{mT}$.
b) ESR spectra produced by the low-energy manifold states of the spin array described in panel a), measured by sensors at different locations (red dots in the inset scheme). The sensors are located at a distance of $0.5 \, \text{nm}$ from the array. In this example, $J=18 \,\text{meV}$, $T = 1 \, \text{K}$, $B_z=1 , \text{T}$, and $\beta = 0.09$.}
\label{fig:2}
\end{figure*}

We now consider that an STM-ESR-active spin picks up the straight field generated by a nearby Haldane spin chain. In ESR-STM experiments the DC current across the STM-surface junction is measured as a function of the frequency of the driving voltage.
The ESR-STM spectrum for this lateral-sensing setup can be described by the following equation \cite{choi17, delCastillo23}:
\begin{eqnarray}
I_{DC}(f)= \sum_{\ell} p_{\ell} {\cal L}(f-f(\ell)),
\label{eq:ESR-STM}
\end{eqnarray}
where the sum runs over the eigenstates of the Hamiltonian (eq. \ref{eq:ham}) of the spin chain, ${\cal H}|\ell\rangle=E_\ell|\ell\rangle$,  
$p_{\ell}=\frac{1}{Z}e^{-E_{\ell}/(k_B T)}$ are
the thermal occupations  of  each eigenstate and ${\cal L}(f-f_{\ell})$ is a Lorentzian type resonance curve centered around the frequency $f_{\ell}$.
We assume that the external magnetic field is perpendicular to the sample so that
the stray field created by the chain at the sensor location is also perpendicular to the substrate.
\footnote{Since both the external field and the exchange interactions are much larger, it is safe to neglect the backaction effect of the stray field of the sensor on the Hamiltonian of the spin chain. }
As a result, the resonant  frequency of the sensor shifts linearly with the stray field: 
\begin{equation}
f(\ell)=\frac{\mu_B g_{\rm s}}{h} (B+b(\ell)),
\label{eq:wsen}
\end{equation}
where 
$h=2\pi \hbar$, and $g_{\rm s}$ is the gyromagnetic factor of the sensor, 
$B$ and $b_{\ell}$ denote the external field and the stray field generated by the target spins in the state $\ell$, respectively, both along the off-plane direction.

 In turn, the stray field 
is given by:
\begin{equation}
    b^z(\ell)  = -\frac{\mu_0}{4\pi}\sum_{n=1}^N\frac{m_n^z(\ell)}{(d^n)^3} , 
\label{eq:stray}
\end{equation}
where $d^n$ is the distance between the sensor and the spin $n$. The magnetic moment vector $m_n^z (l)$ is generated by the spin $n$ in each state $\ell$, and its components are given by:
\begin{equation}
    m_n^z(\ell)= -g\mu_B \langle \ell|S^z_n|\ell\rangle.
\label{eq:mmoments}
\end{equation}
Equations \ref{eq:ESR-STM}, \ref{eq:wsen} \ref{eq:stray}, \ref{eq:mmoments}  relate  the expectation value of the magnetic moments in a given  Haldane-chain state  $|\ell\rangle$ 
to the ESR-STM spectrum of a nearby sensor. 
We now discuss how to determine the magnetic moment of these states and verify fractionalization.  
Importantly, we assume that the temperature is much smaller than the Haldane gap \footnote{We note that, for the case of nanographene Haldane spin chains, the range of temperatures where ESR-STM has been implemented with $T<4 K$, much smaller than as $\Delta_H$ was found to be in the range of 10 meV.}, so that only the four states of the ground state manifold of the chain contribute to the sum in eq. (\ref{eq:ESR-STM}).  Only two of these states, with $S=1$ and $S^z=\pm 1$ have a non-vanishing expectation value of the spins. The corresponding
 magnetic profile of the  $S^z=+1$ states 
calculated using DMRG \cite{white92,white93,catarina23, lado2023dmrgpy}, is shown in figure 1b, for $\beta=0.09$, relevant for triangulene spin chains\cite{Mishra2021,henriques23}. It corresponds to two physically separated objects with $S^z=1/2$ localized at the edges. The $S^z=-1$ has analogous properties. In contrast,  the expectation value of the spin operators is identically zero when calculated with the $S^z=0$ states. Consequently, the four low-energy states of the Haldane spin chain correspond to three distinct magnetic states, with a vanishing stray field for the $S=0$ and $S=1, S^z=0$ states, and a finite stray field of opposite sign for the $S^z=\pm 1$ states (see Fig. (\ref{fig:2}a). As a result, the ESR-STM spectrum of the spin-sensor has three distinct peaks corresponding to three different stray fields (See figure \ref{fig:2}b). Expectedly,  the stray fields of the $S^z=\pm 1$ have the same magnitude and opposite sign.
From the splitting of these peaks, it is possible to pull out the value of the stray field at the location of the sensor:
\begin{equation}
    b^z (\pm1)=\frac{h}{g_s\mu_B} (f_0-f_{\pm 1})
\label{eq:shifts}
\end{equation}
here, $f_{\pm 1}$ represents the resonant frequencies measured at the sensor when the Haldane spin chain occupies the states with $S^z =\pm 1$ within the ground state manifold. 

In order to determine the magnetic moments of the $N/2$ spins of one half of the chain, that define a vector ${\cal M}\equiv(m_1(\pm),...,m_{N/2}(\pm))$, we need to measure the stray field in a set of different locations $N_M$,  that yields a vector in the readout-location space, ${\cal B}\equiv (b_1(\pm),...,b_{N_M}(\pm))$. We have considered two methods to pull out the vector ${\cal M}$ out of ${\cal B}$. The first method is the full inversion of the Biot-Savart's equation (FIBS), that can be written down as  ${\cal{B}}=-(\mu_0/4\pi)\textbf{D}\cal{M}$, where the elements of the matrix \textbf{D} are $|d^n_{n_m}|^{-3}$. In this case, it is apparent that the number of necessary readouts equals half of the chain length, $N_M=N/2$. The second method involves the use of artificial neural networks and requires a dramatically smaller number of measurements to determine the magnetization map (Fig. 3c,d).

The finite magnetic sensitivity of the readouts, denoted by $\delta {\cal{B}}$  imposes 
 an uncertainty in the determination of the edge magnetic moment. 
The sensor spectral resolution is ultimately limited by
the shot noise \cite{supp}\nocite{taylor08,dreau11,barry20}:
\begin{equation}
\delta B_{\text{min}}= 
 \frac{4}{3} 
 \frac{h \delta f}{g_s \mu_B } 
 \sqrt{\frac{e}{I_0 \Delta t}}
\label{sensitivity2}
\end{equation}
where $\delta f $ is the linewidth, $I_0$ is the maximal current in the resonance peak, and $\Delta t$ is measurement time, that may be limited by factors such as thermal drift of the tip.  The associated minimal shift in the sensor frequency is given by $ \Delta f{\rm min}= g\mu_B \delta B_{\rm min}/ h$ in the range of 1MHz have been reported in STM-ESR magnetometry \cite{choi17}.

In the case of the FIBS, the uncertainty of the edge magnetic moments is
given by:
\begin{equation}
    \delta \langle S^z \rangle = \left(\frac{4\pi }{\mu_0\mu_B g } \sum_{n,n_m} |(\textbf{D}^{-1})^n_{n_m}|\right)
     \delta B_{\rm min}.
\end{equation}
\begin{figure}[ht]
\centering
\includegraphics[width=1\linewidth]{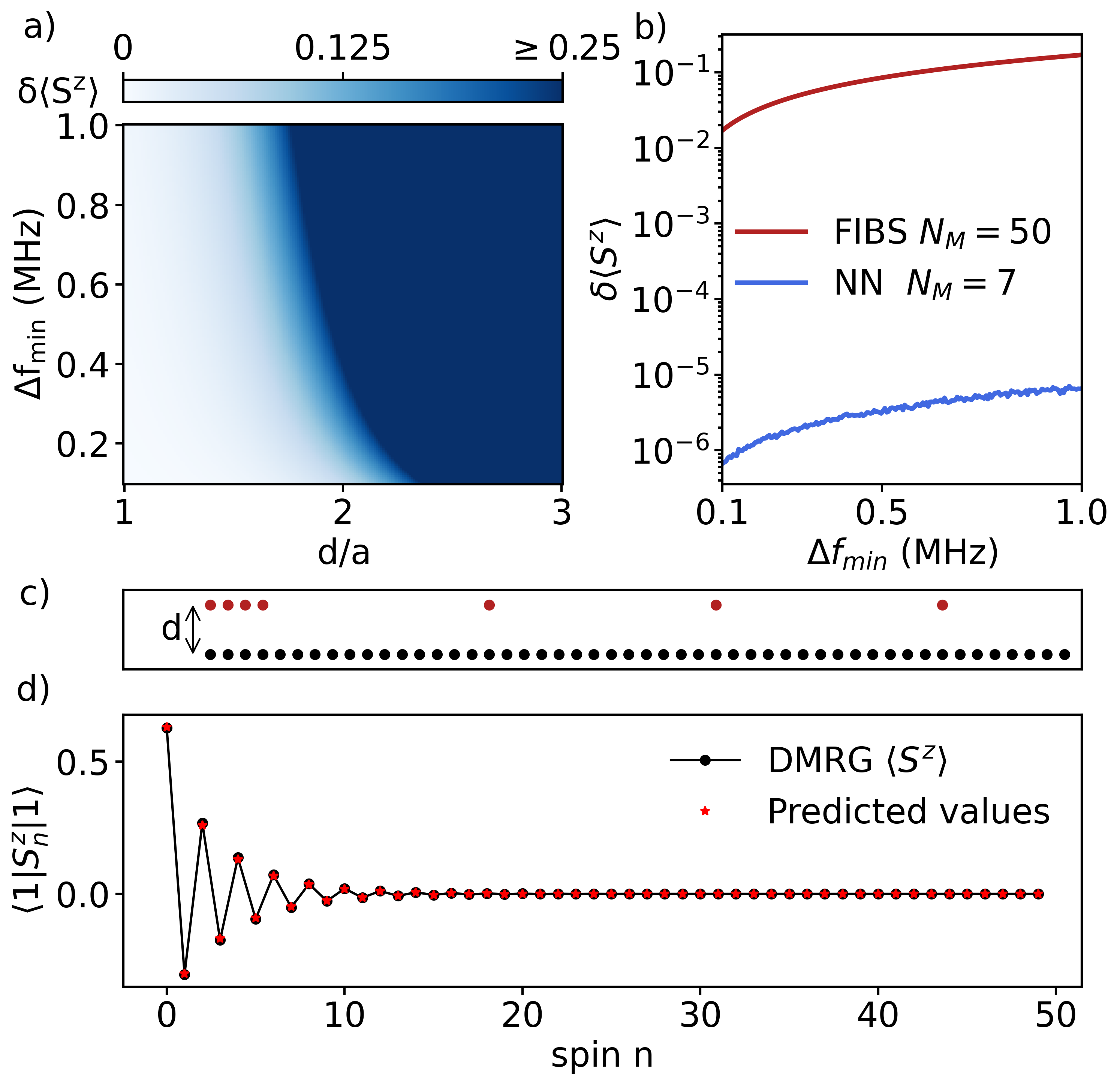}
\caption{ a) $\delta \langle S^z \rangle$ as a function of both, the perpendicular distance of the $N_m$ sensors to the chain (d) and the spectral resolution $\Delta f_{min}$ for a $S=1/2$ sensor, with $N_m=N/2$ different locations from $(x=0,y=d)$ to $(x=+N_m a, y=d)$. b) Error comparison between linear inversion of the Biot-Savart's equation (FIBS) and NN using $N_m=7$ sensor positions ($d=0.5$ nm) c) Location of the sensor positions of the NN configuration (red dots) near the spin chain (black dots) d) Neural network prediction of the expectation values of the spins using a magnetic profile with noise compared to the original values }
\label{fig:3}
\end{figure}

In Figure \ref{fig:3}a, we show that assuming the reported resolution, $\Delta f{\rm min}=1 \, \text{MHz}$ \cite{choi17}),  the uncertainty in the determination of the edge spin, $\delta \langle S^z \rangle$,  is an order of magnitude lower than $\langle S^z \rangle$.
In principle, longer readout times would make it possible to decrease $\delta f$, and thereby, $\delta \langle S^z\rangle$.


We now discuss a second method to invert the Biot-Savart equation that makes use of artificial neuron networks (NN) to invert Biot-Savart's equation. This method comes with two advantages. First, it reduces the number of ESR-STM measurements. Second, it yields a dramatically smaller uncertainty in the determination of the fractional spin. 
 Our approach involves two different NNs. The first NN classifies magnetic profiles derived from ESR readouts, confirming the presence of the
 characteristic Haldane spin chain edge  magnetization (See Fig. (1b)). 
 The second NN is trained to convert Haldane-type magnetic profiles, with a specific number of measurements, into spin expectation values. The training set incorporates several thousand spin distributions obtained varying $\beta$  between $0$ and $1/3$, each with its characteristic magnetic profile. In addition, we added random noise with amplitude bounded by the magnetic sensitivity range (See Supplemental Material \cite{supp}). The synergistic application of these two neural networks reduces the training time to a few minutes on a conventional laptop. This is crucial since a distinct NN needs to be trained for each experimental layout, considering factors such as spin arrangement, sensor positions, sensor type, and spectral resolution.

In  Figure \ref{fig:3}b, we used $N_M=7$ strategically positioned measurements \cite{supp}, as shown in Figure \ref{fig:3}c. Our calculations reveal that, with a number  ESR readouts much smaller than $N/2$, it is possible to obtain the magnetization maps and, therefore, the total magnetic moment of half of the chain with an uncertainty as low as $\delta \langle M_z \rangle \approx 10^{-4}\mu_B$ (See Supplemental Material \cite{supp}), as shown in Figure \ref{fig:3}b. Consequently, this methodology proves sufficient for determining the presence of an $S=1/2$ object and its localization length, using state-of-the-art ESR-STM instrumentation.

We now discuss how to infer another important property of the edge spins, namely, their localization  
length $\xi$. This is based on two facts. 
First,  the inversion of the Biot-Savart equation yields the value of the magnetic moment at many sites close to the edge. Second, our numeric work shows that we can parametrize the spins with the following equation:
\begin{equation}
\langle S_n^z\rangle_{\pm}= \pm  (-1)^n {\cal A}e^{-\frac{n}{\xi}}
\label{eq:edgespin}
\end{equation}
where  $\mathcal{A}$ represents the maximum value of $\langle S^z_n \rangle$ at the edge spin, ensuring that $\sum_{n=1}^{N/2} \langle S^z_n \rangle = \frac{1}{2}$. Our numerical calculation (see Supplemental Material \cite{supp})
shows that the second moment of the magnetization field 
$\langle n \rangle = \frac{\sum_n |\langle S^z_n \rangle| \cdot n}{\sum_n |\langle S^z_n \rangle|}
$ 
is proportional to $\xi$ in an almost one-to-one relation ($\langle n \rangle \approx 1.02 \, \xi$), which permits one to determine this quantity 
with an uncertainty associated to $\delta \langle S_n^z \rangle$. For the reported spectral resolution of 1 MHz and d=0.5 nm, the relative errors would be 
$\frac{\delta \xi}{\xi} \approx 10^{-1}  $ for FIBS and $\frac{\delta \xi}{\xi} \approx 10^{-5}$ for NNs \cite{supp}.

In conclusion, we propose a method to measure the two outstanding properties of the $S=1/2$ fractional degrees of freedom that emerge at the edges of Haldane $S=1$ chains: their fractional magnetic moment $M=g\mu_B S$ and their localization length or spatial extension, $\xi$. Our theoretical analysis shows that our method can be implemented with state-of-the-art ESR-STM magnetometry.
Our proposal permits one to go beyond previous work, where the presence of fractional degrees of freedom is inferred indirectly, but the actual fractionalization of the magnetic moment is not measured directly. This approach could be used to probe fractional edge spins expected to occur in 
two-dimensional AKLT models 
and could also inspire similar experiments using related atomic scale magnetometers, such as NV centers \cite{grinolds13,rondin14,chatterjee19,finco23}. 

We acknowledge Arzhang Ardavan for fruitful discussions and Jose Lado for technical assistance on the implementation of DMRG.
J.F.R.  acknowledges financial support from 
FCT (Grant No. PTDC/FIS-MAC/2045/2021),
SNF Sinergia (Grant Pimag),
Generalitat Valenciana funding Prometeo2021/017
and MFA/2022/045, and funding from MICIIN-Spain (Grant No. PID2019-109539GB-C41).
YDC acknowledges funding from FCT, QPI,  (Grant No.SFRH/BD/151311/2021) and thanks the hospitality of the Departamento de F\'isica Aplicada at the Universidad de Alicante.

\bibliographystyle{apsrev4-2}
\bibliography{biblio}{}

\end{document}


\title{Supplemental material of "Probing spin fractionalization with ESR-STM absolute magnetometry" }

\author{Y. del Castillo$^{1,2}$, J. Fern\'{a}ndez-Rossier$^1$\footnote{On permanent leave from Departamento de F\'{i}sica Aplicada, Universidad de Alicante, 03690 San Vicente del Raspeig, Spain}$^,$\footnote{joaquin.fernandez-rossier@inl.int} }

\affiliation{$^1$International Iberian Nanotechnology Laboratory (INL), Av. Mestre Jos\'{e} Veiga, 4715-330 Braga, Portugal }
\affiliation{$^2$Centro de F\'{i}sica das Universidades do Minho e do Porto, Universidade do Minho, Campus de Gualtar, 4710-057 Braga, Portugal }

\date{\today}

    
\maketitle
\section{Shot-noise limited spectral resolution}
Here we provide a derivation of the limits that electron shot-noise imposes on the spectral sensitivity of ESR-STM and derive eq (8) in the manuscript. Our derivation follows a closely similar analysis to optically detected magnetic resonance \cite{dreau11,taylor08,barry20}, where photon shot-noise sets the limit of spectral resolution.
There are two  key ideas. First, any variation of the resonance frequency, ultimately coming from variations in the magnetic field,  produces a variation in current given by: 
\begin{equation}
    \Delta I = \frac{\partial I }{\partial f} \Delta f
\end{equation}
The second key idea is that shot noise imposes a lower threshold to the minimum current variation that can be detected.
Let us define $\Delta I|_{\rm min}$ as the minimal current change that can be measured in a given set-up, and $\frac{\partial I}{\partial f}|_{\rm max}$ as the maximal variation of current with respect to frequency. We can thus write an expression for the smallest variation of frequency $\Delta f|_{\rm min}$ that could be measured in such system:
\begin{equation}
\Delta f|_{\rm min}= 
 \frac{1}{\frac{\partial I}{\partial f}|_{\rm max}}
 \Delta I|_{\rm min} 
 \label{eq:main}
\end{equation}

We now obtain an expression for $\Delta f|_{\rm min}$ assuming that the minimal current variation that can be recorded is shot-noise limited, $\Delta I|_{\rm min}>\delta I_{\rm shot \, noise}$. The shot noise current can be obtained as follows. The average current $I$ relates to the average number of electrons $N$ with charge $e$ that goes through the junction in a time interval $\Delta t$:
 $I=\frac{eN}{\Delta t}$.
Expressed in terms of the number of electrons, the shot noise is given by the Poissonian distribution, $\sqrt N$:
\begin{equation}
    \delta I= \sqrt{\langle (I-\langle I\rangle)^2\rangle}=\frac{e\sqrt{N}}{\delta t}= \sqrt{\frac{e I}{\Delta t}}
\end{equation}
Hence, we can now write:
\begin{equation}
\Delta f|_{\rm min}\geq 
 \frac{1}{   \frac{\partial I}{\partial f}|_{\rm max}} \sqrt{\frac{e I}{\Delta t}}
 \label{eq:main2}
\end{equation}
Assuming a given functional dependence of DC current on frequency, it is possible to derive an expression for
$\frac{\partial I}{\partial f}|_{\rm max}$ in terms of the spectral width of the resonance,  $\delta f$.  In our case, we assume a Lorentzian resonance curve and we obtain the spectral sensitivity in terms of both  $\delta f$ and the measurement time:
\begin{equation}
\Delta f|_{\text{min}}= 
  \frac{4}{3 }  \delta f 
 \sqrt{\frac{e}{ I_0\Delta t}}
 =
  \frac{4 }{3}\delta f
 \sqrt{\frac{1}{N}}
\label{sensitivity2}
\end{equation}
where, in the last step, we express the results in terms of  $N$, the number of tunneling electrons during the measurement time.

\section{Full Inversion of Biot-Savart's equation (FIBS)}
 
The stray magnetic field generated by $N$ spins at a specific location is described by the dipolar formula:

\begin{equation}
    \vec b (\vec{r}, \ell) = \frac{\mu_0}{4\pi} \sum_{n=1}^N \left [   \frac{3 (\vec{m_n}(\ell) \cdot \vec{r_n})\vec{r_n} }{|d_n|^5} - \frac{\vec{m_n }(\ell)}{|d_n|^3} \right ] 
\label{eq:SMdipolar_formula}
\end{equation}

Here, $\vec{r_n}$ is the vector distance between position $\vec{r}$ and spin $n$, $\vec{m_n}$ is the magnetic moment vector of spin $n$, and $d_n=\vec r -\vec{r^n}$. Our assumption is that the magnetic moments have only a non-zero component in the off-plane direction ($z$), this implies $\vec{m_n}(\ell) \cdot \vec{r_n}=0$. Consequently, the stray field simplifies to:

\begin{equation}
   b_z (\ell) = -\frac{\mu_0}{4\pi} \sum_{n=1}^N \frac{m_n^z (\ell)}{|d_n|^3} 
\label{eq:SMdipolar}
\end{equation}

When the sensor is placed at different locations, yielding $N_M$ different stray fields, we express this as a matrix equation ${\cal B}=-\frac{\mu_0}{4 \pi}\textbf{D}{\cal M}$:

\begin{equation}
\begin{pmatrix}
    b^{1}_z \\ \vdots \\ b^{N_M}_z
\end{pmatrix}
=
-\frac{ \mu_0}{4 \pi}
\begin{pmatrix}
|d_1^1|^{-3} & \cdots & |d^{1}_N|^{-3} \\
\vdots &  \ddots & \vdots \\
|d^{N_M}_1|^{-3} & \cdots & |d^{N_M}_N|^{-3}
\end{pmatrix}
\begin{pmatrix}
     m_1^z \\ \vdots \\ m_N^z
\end{pmatrix}
\end{equation}

A crucial condition for a full inversion is $N_M=N$, ensuring $\textbf{D}$ is a square matrix.

In order to analyze the uncertainties introduced by experimental errors related to the magnetic field, we can use the relation $\delta b = \frac{h}{g_s \mu_B} \Delta f_{min}$, where $\Delta f_{min}$ is the minimal frequency shift that can be detected.  Considering magnetic moments defined as $m^z_n (\ell) = -g\mu_b \langle \ell|S^z_n |\ell\rangle$, we find:

\begin{equation}
    \Delta f_{min} = \frac{\mu_0 \mu_B^2 g_s g }{4 \pi h } \sum_{n=1}^N \frac{\delta \langle S^z_n \rangle}{|d_n|^3} 
\end{equation}

After $N_M$ measurements, the errors are given by:

\begin{equation}
\begin{split}
\begin{pmatrix}
    \delta \langle S_1^z\rangle  \\ \vdots \\ \delta \langle S_N^z\rangle
\end{pmatrix}
= &
\frac{\mu_0 \mu_B^2 g_s g }{4 \pi h }
\begin{pmatrix}
|d_1^1|^{-3} & \cdots & |d^{1}_N|^{-3} \\
\vdots &  \ddots & \vdots \\
|d^{N_M}_1|^{-3} & \cdots & |d^{N_M}_N|^{-3}
\end{pmatrix}^{-1}
\begin{pmatrix}
    \Delta f_{min} \\ \vdots \\ \Delta f_{min}
\end{pmatrix} \\
=& \frac{\mu_B^2g_sg}{h}\textbf{D}^{-1}\begin{pmatrix}
    \Delta f_{min} \\ \vdots \\ \Delta f_{min}
\end{pmatrix} \\
\end{split}
\end{equation}

Assuming independence of the measurements and identical $\Delta f_{min}$ for each, the total error can be obtained as follows:

\begin{equation}
    \delta \langle S^z \rangle = \sum_{n=1}^{N} |\delta \langle S^z_n \rangle| = \Delta f_{min} \frac{\mu_B^2g_sg}{h}  \sum^{N;N_M}_{n=1;n_m=1} |(\textbf{D}^{-1})^{n_m}_{n}|
\label{eq:total_error}
\end{equation}

\section{ Neural network inversion of Biot-Savart's equation }

The full inversion of the Biot-Savart's equation has two drawbacks. First, it requires performing a number of ESR-STM readouts
of the order of half of the Haldane chain size. Second, derived from the first, the error associated with 
 \ref{eq:total_error} scales up with the number of readouts. Here we discuss the technical aspects of an alternative method, that makes use of two neural networks (NN).

\begin{itemize}
    \item Classification NN: After the measurements, we will have a magnetic profile of the system. This first categorical network classifies between two groups of profiles, depending on whether or not they are compatible with the edge magnetization profile of the Haldane spin chain model. Other magnetic profiles generated by, for example, antiferromagnetic, ferromagnetic, or random systems in the same spin locations, are included in the training set.
    \item Expectation Values NN: Only if the classification in the first NN is affirmative, so we know the system matches the characteristics of the training set (Haldane spin chains), the second NN predicts the expectation values of the spins. The training set is composed of thousands of spin densities of Haldane spin chains with different $\beta$ values and their associated magnetic profiles generated with the addition of random noise bounded by the spectral resolution. 
\end{itemize}

The locations of the sensor and the Haldane spin chain and the spectral resolution are predetermined by the experiments. Then, the neural networks are trained to align with those specifications. By using the first classification NN, the second one can focus only on Haldane spin densities. This allows us to build much simpler and easier-to-train NNs. 

\subsection{Guidelines for the choice of STM-ESR readout grid }

The STM-ESR readout grid is defined by the set of points where the probe atom is placed to map out the stray field of the chains. Here we provide some guidelines to optimize their placement for maximum information capture.

Considering that the number of sensors in the NNs approach is smaller than the number of spins, placing them on the edge spins is ideal since the spin 1/2 quasiparticle is mostly located there. However, this could result in a lack of information on the center of the chain since any magnetic detail beyond a certain distance would be below the spectral resolution. We could not completely rely on the classification results of the first NN.

We conclude that the optimal sensor placement for a Haldane spin chain involves situating as many sensors as possible near the edge spins and additionally, a few sensors along the chain to confirm the characteristic behavior of a Haldane chain. An example of this arrangement is shown in Figure 3c in the manuscript with the sensors as red dots. 

\subsection{Errors}
\begin{figure}[h]
    \centering    \includegraphics[width=0.33\linewidth]{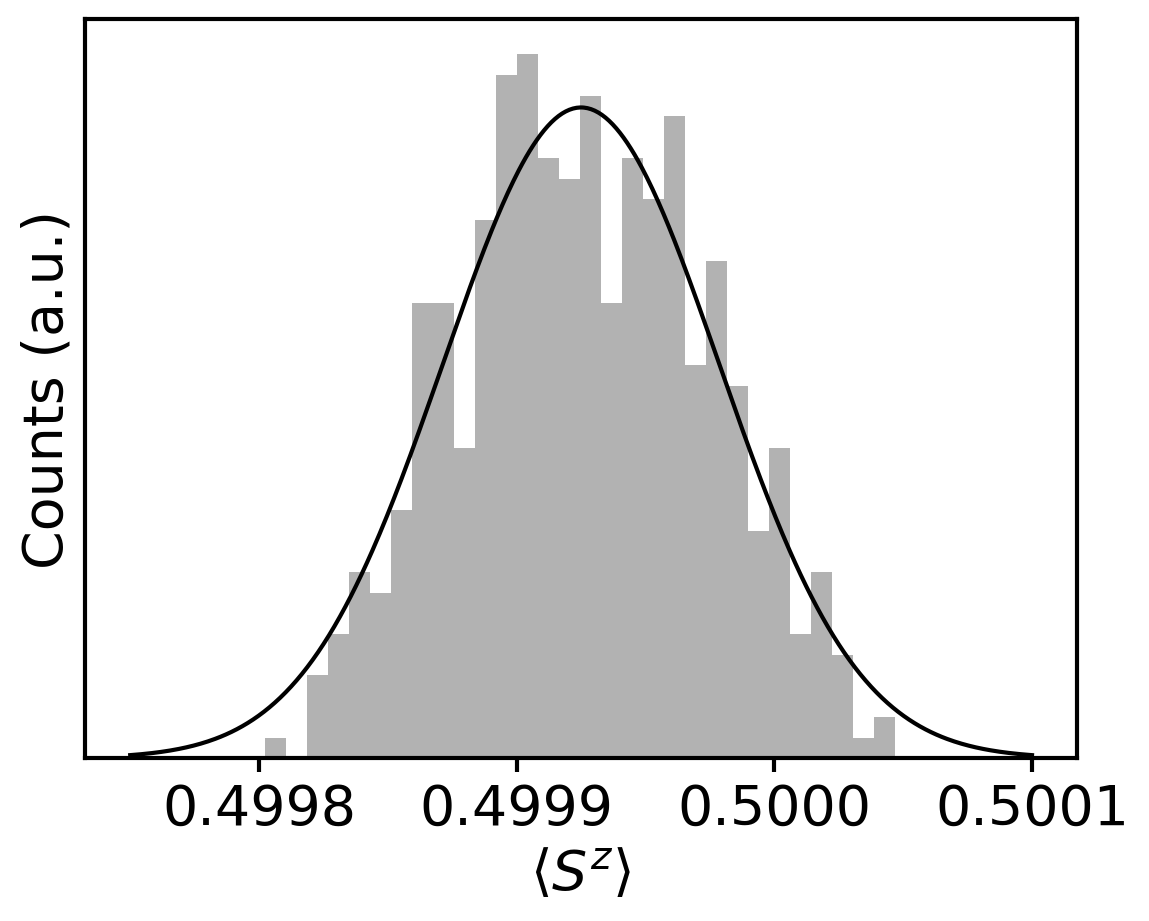}
    \caption{Gaussian fit on the density plot of $\langle S^z \rangle$ for each prediction using random noise (within the range of the spectral resolution) for the same Haldane chain. $d=0.5$ nm, $a=0.3$ nm, $\Delta f_{min}=1$ MHz, $N_M=7$.}
    \label{fig:ERROR-NN}
\end{figure}

The error analysis needs to be performed for each NN, so the conditions match those of the experiment. To carry out the analysis, first, we take the spin density of a Haldane chain. Second, we compute the stray field on the predefined sensor positions of the NN and we generate hundreds of sets of \textit{measurements} by introducing random errors within the range of the spectral resolution. We use the NN to predict the value of the spins to sum them and finally, we obtain $\langle S^z \rangle$. Then, we can analyze the standard deviation of all the predictions as shown in Figure \ref{fig:ERROR-NN}. In this particular example, the NN has an uncertainty of around $\delta \langle S^z \rangle /\langle S^z \rangle =10^{-4}$ for predictions using test sets with random noise bounded by a spectral resolution of $\Delta f_{min}\approx 1\,  \rm MHz$. This process is repeated for different spectral resolutions, enabling us to compare with Equation \ref{eq:total_error}. We show this comparison, for a particular sensor arrangement, in Figure 3b in the manuscript.

\section{Determination of the edge spin localization length $\xi$}

\begin{figure}[h!]
    \centering    \includegraphics[width=0.6\linewidth]{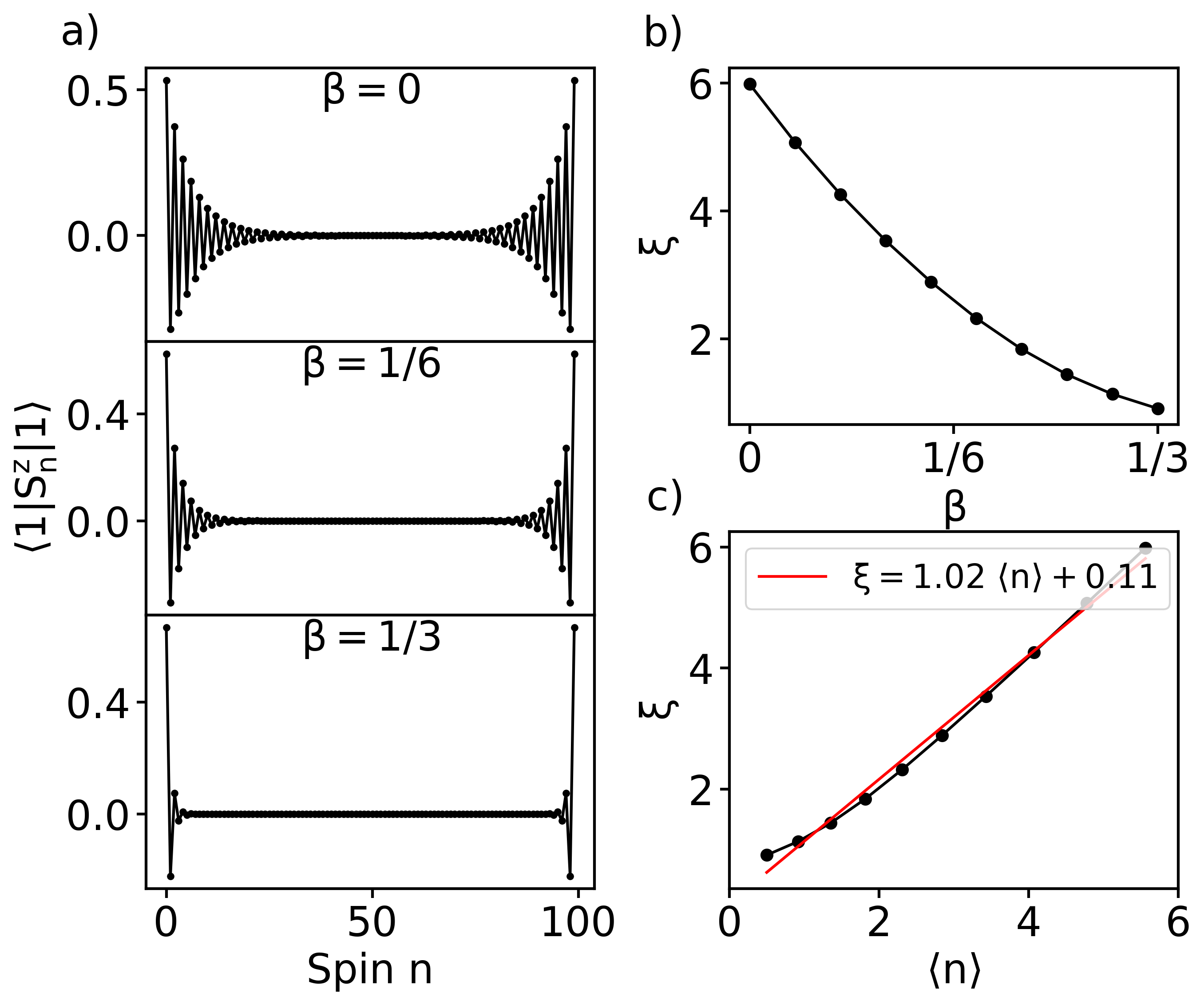}
    \caption{a) Spin density of a N=100 Haldane spin chain for three different $\beta$ values. b) Localization length obtained from the exponential fitting for half of the chain as a function of $\beta$. c) The second moment of the spin distribution computed for the same systems and compared with the correlation length. In red, a lineal fit shows an almost one-to-one relation. }
    \label{fig:position-op}
\end{figure}

We fit several spin densities of half Haldane spin chain with different $\beta$ values (see for example Figure \ref{fig:position-op}a) to the function $\langle \ell |S^z_n| \ell \rangle= (-1)^n {\cal A} (\ell) e^{-n/\xi}$. From the fit we can obtain $\xi$, what we define as as the localization length. We plot $\xi$ as a function of $\beta$ in Figure \ref{fig:position-op}b. As shown in the manuscript, the second moment of the spin distribution is related to the localization length in an almost one-to-one relation:
\begin{equation}
    \langle n \rangle = \frac{\sum_{n=1}^{N/2}| \langle \ell |S^z_n| \ell \rangle | \cdot n}{\sum_{n=1}^{N/2} |\langle \ell |S^z_n| \ell \rangle |}
\label{eq:n_op}
\end{equation}
where we sum over $N/2$ so we only consider half of the chain. In Figure \ref{fig:position-op}c, using a linear fit, we show that, aside from an offset of $0.11$, the relation is almost one-to-one ($1.02$).

\subsection{Errors}
We need to consider the error of each individual spin to compute the uncertainties of the localization length. We use equation \ref{eq:total_error}, but we sum only over the sensor read-out space. Thus, the error of each spin value can be defined by the following expression:

\begin{equation}
    \delta \langle S^z_n \rangle  = \Delta f_{min} \frac{\mu_B^2g_sg}{h}  \sum^{N_M}_{n_m=1} |(\textbf{D}^{-1})^{n_m}_{n}|
\end{equation}

Now we can calculate the uncertainty of the correlation length using the second moment of the spin distribution:
\begin{equation}
\begin{split}
        \delta \langle n\rangle = & \sum_{i=1}^{N/2} \left ( \frac{\partial \langle n \rangle}{\partial \langle S^z_i \rangle} \delta \langle S^z_i \rangle \right )= \sum_{i=1}^{N/2} \left (\frac{i\sum_n |\langle S^z_n \rangle| - \sum_n |\langle S^z_n\rangle| \cdot n }{(\sum_n |\langle S^z_n \rangle|)^2}\delta \langle S^z_i \rangle \right )\\
        = & \sum_{i=1}^{N/2}  \left ( \frac{i - \langle n \rangle}{\sum_n |\langle S^z_n \rangle|} \delta \langle S^z_i \rangle \right )\\
\end{split}
\end{equation}

Finally, we illustrate the error analysis with a numerical example. Using FIBS, we take the following conditions: $\Delta f_{min}=1 \, \rm MHz$,$N_m=N/2$ for a $N=100$ chain located at $0.45 \, \rm nm$ for each spin with a spacing of $0.3 nm$.  We can take the value from the already computed uncertainty of the spin value of Figure 3a in the manuscript for those conditions,  $\delta \langle S^z \rangle \approx 0.05$ so for each spin will be $\delta \langle S^z_n \rangle \approx 0.075/N_m$. The result shows that the relative error is $\delta \langle n\rangle/\langle n\rangle \approx 10^{-1}$. Now, we compare it with the NNs approach. Using $\delta \langle S^z_n \rangle \approx 10^{-5}/N_m$ shown in Figure 3b from the manuscript for that particular distance, we obtain the following relative uncertainty $\delta \langle n\rangle/\langle n\rangle \approx 10^{-5}$.

\bibliographystyle{apsrev4-2}
\bibliography{biblio}{}